
\frenchspacing

\parindent15pt
\overfullrule=0pt
\abovedisplayskip4pt plus2pt
\belowdisplayskip4pt plus2pt 
\abovedisplayshortskip2pt plus2pt 
\belowdisplayshortskip2pt plus2pt  

\font\twbf=cmbx10 at12pt
 at12pt
 at12pt

\font\sc=cmcsc10

\font\ninerm=cmr9 
\font\nineit=cmti9 
\font\ninesy=cmsy9 
\font\ninei=cmmi9 
\font\ninebf=cmbx9 

\font\sevenrm=cmr7  
 
\font\seveni=cmmi7  
\font\sevensy=cmsy7 

\font\fivenrm=cmr5  
\font\fiveni=cmmi5  
\font\fivensy=cmsy5 

\def\nine{%
\textfont0=\ninerm \scriptfont0=\sevenrm \scriptscriptfont0=\fivenrm
\textfont1=\ninei \scriptfont1=\seveni \scriptscriptfont1=\fiveni
\textfont2=\ninesy \scriptfont2=\sevensy \scriptscriptfont2=\fivensy
\textfont3=\tenex \scriptfont3=\tenex \scriptscriptfont3=\tenex
\def\rm{\fam0\ninerm}%
\textfont\itfam=\nineit    
\def\it{\fam\itfam\nineit}%
\textfont\bffam=\ninebf 
\def\bf{\fam\bffam\ninebf}%
\normalbaselineskip=11pt
\setbox\strutbox=\hbox{\vrule height8pt depth3pt width0pt}%
\normalbaselines\rm}

\hsize30cc
\vsize44cc
\nopagenumbers

\def\luz#1{\luzno#1?}
\def\luzno#1{\ifx#1?\let\next=\relax\yyy
\else \let\next=\luzno#1\xxx\fi\next}
\def\sp#1{\def\xxx{\kern1.7pt}\def\yyy{\kern-1.7pt}\luz{#1}}
\def\spa#1{\def\xxx{\kern1pt}\def\yyy{\kern-1pt}\luz{#1}}

\newcount\beg
\newbox\aabox
\newbox\atbox
\newbox\fpbox
\def\abbrevauthors#1{\setbox\aabox=\hbox{\sevenrm\uppercase{#1}}}
\def\abbrevtitle#1{\setbox\atbox=\hbox{\sevenrm\uppercase{#1}}}
\long\def\pag{\beg=\pageno
\def\leftheadline{\noindent\rlap{\nine\folio}\hfil\copy\aabox\hfil}
\def\rightheadline{\noindent\hfill\copy\atbox\hfill\llap{\nine\folio}}
\def\phead{
\noindent\vbox{
\centerline{\hfill LMU-TPW 96-6\ }

\centerline{\hfill February 8, 1996\ }}}
\footline{\ifnum\beg=\pageno \hfill\nine[\folio]\hfill\fi}
\headline{\ifnum\beg=\pageno\phead
\else
\ifodd\pageno\rightheadline \else \leftheadline \fi 
\fi}}

\newbox\tbox
\newbox\aubox
\newbox\adbox
\newbox\mathbox

\def\title#1{\setbox\tbox=\hbox{\let\\=\cr 
\baselineskip14pt\vbox{\twbf\tabskip 0pt plus15cc
\halign to\hsize{\hfil\ignorespaces \uppercase{##}\hfil\cr#1\cr}}}}

\newbox\abbox
\setbox\abbox=\vbox{\vglue18pt}

\def\author#1{\setbox\aubox=\hbox{\let\\=\cr 
\nine\baselineskip12pt\vbox{\tabskip 0pt plus15cc
\halign to\hsize{\hfil\ignorespaces \uppercase{\spa{##}}\hfil\cr#1\cr}}}%
\global\setbox\abbox=\vbox{\unvbox\abbox\box\aubox\vskip8pt}}

\def\address#1{\setbox\adbox=\hbox{\let\\=\cr 
\nine\baselineskip12pt\vbox{\it\tabskip 0pt plus15cc
\halign to\hsize{\hfil\ignorespaces {##}\hfil\cr#1\cr}}}%
\global\setbox\abbox=\vbox{\unvbox\abbox\box\adbox\vskip16pt}}

\def\mathclass#1{\setbox\mathbox=\hbox{\footnote{}{1991 {\it Mathematics 
Subject Classification}\/: #1}}}

\long\def\maketitlebcp{\pag\unhbox\mathbox
\vglue7cc
\box\tbox
\box\abbox
\vskip8pt}

\long\def\abstract#1{{\nine{\bf Abstract.} 
#1

}}

\def\section#1{\vskip-\lastskip\vskip12pt plus2pt minus2pt
{\bf #1}}

\long\def\th#1#2#3{\vskip-\lastskip\vskip4pt plus2pt
{\sc #1} #2\hskip-\lastskip\ {\it #3}\vskip-\lastskip\vskip4pt plus2pt}

\long\def\defin#1#2{\vskip-\lastskip\vskip4pt plus2pt
{\sc #1} #2 \vskip-\lastskip\vskip4pt plus2pt}

\long\def\remar#1#2{\vskip-\lastskip\vskip4pt plus2pt
\sp{#1} #2\vskip-\lastskip\vskip4pt plus2pt}

\def\Proof{\vskip-\lastskip\vskip4pt plus2pt 
\sp{Proo{f.}\ }\ignorespaces}

\def\endproof{\nobreak\kern5pt\nobreak\vrule height4pt width4pt depth0pt
\vskip4pt plus2pt}

\newbox\refbox
\newdimen\refwidth
\long\def\references#1#2{{\nine
\setbox\refbox=\hbox{\nine[#1]}\refwidth\wd\refbox\advance\refwidth by 12pt%
\def\textindent##1{\indent\llap{##1\hskip12pt}\ignorespaces}
\vskip24pt plus4pt minus4pt
\centerline{\bf References}
\vskip12pt plus2pt minus2pt
\parindent=\refwidth
#2

}}

\def\footnoterule{\kern -3pt \hrule width 4cc \kern 2.6pt}

\catcode`@=11
\def\vfootnote#1%
{\insert\footins\bgroup\nine\interlinepenalty\interfootnotelinepenalty%
\splittopskip\ht\strutbox\splitmaxdepth\dp\strutbox\floatingpenalty\@MM%
\leftskip\z@skip\rightskip\z@skip\spaceskip\z@skip\xspaceskip\z@skip%
\textindent{#1}\footstrut\futurelet\next\fo@t}
\catcode`@=12

\def\ra{\rangle}
\def\la{\langle}

\def\zet{\hbox{\bf Z}}

\def\hub{Hubbard model}

\def\klu{{K_l^{(+)}}}
\def\kld{{K_l^{(-)}}}
\def\klz{{K_l^{(z)}}}

\def\kz{{K^{(z)}}}
\def\krz{{K_r^{(z)}}}
\def\ku{{K^{(+)}}}
\def\kd{{K^{(-)}}}
\def\kjz{{K_j^{(z)}}}
\def\kiz{{K_i^{(z)}}}

\def\aiu{{a_{i\uparrow}}}
\def\aid{{a_{i\downarrow}}}
\def\aaiu{{a^{\dagger}_{i\uparrow}}}
\def\aaid{{a^{\dagger}_{i\downarrow}}}
\def\alu{{a_{l\uparrow}}}
\def\ald{{a_{l\downarrow}}}
\def\aalu{{a^{\dagger}_{l\uparrow}}}
\def\aald{{a^{\dagger}_{l\downarrow}}}

\def\nld{{n_{l\downarrow}}}
\def\nlu{{n_{l\uparrow}}}

\def\aju{{a_{j\uparrow}}}
\def\ajd{{a_{j\downarrow}}}
\def\aaju{{a^{\dagger}_{j\uparrow}}}
\def\aajd{{a^{\dagger}_{j\downarrow}}}

\def\nid{{n_{i\downarrow}}}

\def\niu{{n_{i\uparrow}}}

\def\ais{{a_{i\sigma}}}

\def\aais{{a^{\dagger}_{i\sigma}}}

\def\aajs{{a^{\dagger}_{j\sigma}}}

\def\nis{{n_{i\sigma}}}

\def\ei{{e^{-i(\vec{\Phi}-\vec{\kappa}) \cdot 
\vec p_i-i \vec{\kappa} \cdot \vec p_j+\zeta \vec R_{ij} 
   \cdot (\vec x_i-\vec x_j)}}}

\def\epi{{e^{{1 \over 2} \vec R_{ij} \cdot  \vec{\Phi} \zeta \hbar}}}
\def\epj{{e^{-{1 \over 2} \vec R_{ij} \cdot \vec{\Phi} \zeta \hbar}}}

\def\eaijm{e^{-{1 \over 2}\alpha}}
\def\eajim{e^{{1 \over 2}\alpha^*}}

\def\hh{{H_{Hub}}}

\mathclass{Primary 81R50; Secondary 82D55, 82B20.}

\abbrevauthors{B.L. Cerchiai and P. Schupp}
\abbrevtitle{Extended Hubbard Model...}

\title{Symmetries of an Extended Hubbard Model}
\author{Bianca \ Letizia \ Cerchiai}
\address{Sektion Physik, Universit\"at M\"unchen, LS Prof. Wess, \\
Theresienstr. 37, 80333 M\"unchen, Germany\\
E-mail: Bianca.Cerchiai@Physik.Uni-Muenchen.de}

\author{Peter \ Schupp}
\address{Sektion Physik, Universit\"at M\"unchen, LS Prof. Wess, \\
Theresienstr. 37, 80333 M\"unchen, Germany\\
E-mail: Peter.Schupp@Physik.Uni-Muenchen.de}

\maketitlebcp

\abstract{An extended Hubbard model with phonons
is considered on a $D$-dimensional lattice. The symmetries of the model are
studied in various cases. It is shown that for a certain choice of the parameters
a superconducting $SU_q(2)$ holds as a true quantum symmetry---but
only for $D=1$.
}

\section{1. Introduction}

In this article we are going to study extensions of the
{\hub} on a $D$-dimensional lattice and their symmetries.

The {\hub} was originally introduced in [1] and is the
simplest model describing systems of itinerant interacting
electrons in solid-state physics. Its importance is
mainly due to the possibility that it 
may describe high-temperature superconductivity. 

Since the work of Yang and Zhang [2,3] it has been
known that the {\hub} has a $SU(2) \times SU(2))/\zet_2$-symmetry.
This symmetry is the product of two separate $SU(2)$ symmetries: 
a {\it magnetic\/} 
symmetry, which accounts for the (antiferro-)magnetic properties
of the electron system, and a {\it superconductive\/} symmetry, which is
supposed to give rise to superconductivity when it is broken. 

A.\ Montorsi and M.\ Rasetti investigated whether the symmetry of
the standard {\hub} generalizes to a quantum group symmetry
of the {\hub} with phonons. 
In [4] they claim the existence of a {\it superconductive\/}
$SU_q(2)$-quantum group symmetry and a standard {\it magnetic\/}
$SU(2)$-symmetry for a {\hub} with non-local phonon interaction.
We were able to verify this symmetry for a particular extension of the {\hub},
but find it to be restricted to 1-dimensional lattices. Moreover, 
this Quantum Symmetric Hubbard Model is 
not the standard Hubbard Model with Phonons.
   
\section{2. The Extended Hubbard Model} 

The Hamiltonian of the extended {\hub} is given by [4,5]:
$$\eqalignno{
& \hh=H_{el}^{(loc)}+H_{ph}^{(loc)}+H_{el-ph}^{(loc)}+H^{(non-loc)}
& (1) \cr }
$$
where
$$\eqalignno{
& H_{el}^{(loc)}=u \sum_i \niu \nid-\mu \sum_{i,\sigma} \nis & (2) \cr
& H_{ph}^{(loc)}=\sum_i \left({\vec p_i{}^2 \over 2M} 
+{{1} \over {2}} M \omega^2 \vec x_i{}^2\right) & (3) \cr
& H_{el-ph}^{(loc)}=-\vec{\lambda} \cdot \sum_i (\niu + \nid) \vec{x_i}
& (4) \cr
& H^{(non-loc)}=\sum_{\la i,j \ra} \sum_{\sigma}
T_{ij} a^{\dagger}_{j \sigma} a_{i \sigma} & (5) \cr }
$$
with 
$T_{ij} = T^\dagger_{ji}$ given by
$$\eqalignno{
& T_{ij}=t  e^{-\hbar \zeta \vec R_{ij} \cdot \vec \kappa} \exp\{\zeta \vec R_{ij} 
\cdot (\vec x_i-\vec x_j)+i\vec{\kappa} \cdot(\vec p_i-\vec p_j)\} 
& (6) \cr }
$$
In these expressions the Hamiltonian depends on the operators
$\ais,\aais,\vec p_i, \vec x_i$, where
$\ais,\aais$ are fermionic annihilation/creation operators
for an electron with spin $\sigma=\{\uparrow,\downarrow\}$ at site $i$,
while $\vec p_i, \vec x_i$ are momentum
and displacement operators for an ion at site $i$ with the usual
commutation relations.

The Hamiltonian also depends on
$\vec R_i$ which gives the rest position of the the ion at site $i$.
Note that the expression 
$\vec R_{ij} \equiv {(\vec R_j-\vec R_i) \over | \vec R_i-\vec R_j |}$
always has 
unit modulus and that in the one-dimensional case it just amounts to a sign.
$| \vec R_i- \vec R_j |$ is the interatomic distance at equilibrium so that 
it does not depend on $i,j$.
In the following we will explain the various terms composing {$\hh$}:

$H_{el}^{(loc)}$ is the local electron-electron interaction and
is the sum of two terms. The first one, the on-site repulsion
$u \sum_i \niu \nid$,  is reminiscent of the
Coulomb repulsion between the electrons and is determined by
the parameter $u$.
The second one, the chemical potential $-\mu \sum_{i,\sigma} \nis$,
shows that the Hamiltonian is written in a grand-canonical formalism;
the parameter $\mu$ fixes the avarage number of electrons in the
lattice.

$H_{ph}^{(loc)}$ is the kinetic term for the phonons, which  are 
described by a set of decoupled Einstein oscillators, all with
the same frequency $\omega$ and the same mass $M$.

$H_{el-ph}^{(loc)}$ is the local phonon-electron interaction term [6,7] 
with coupling constant $\vec \lambda$,
which at each site gives an attractive force proportional to the number of
electrons and to the ion displacement.

The crucial term is the non-local one. It is a hopping amplitude,
and gives the probability that  an electron can jump from one site
to another one. Notice that we have retained 
only the nearest neighbour terms $\langle ij\rangle$ and 
hence assumed negligible overlap between all other atomic
orbitals. In the extended model this amplitude
depends on the ion displacement and momentum, resulting in
a non-local interaction between the phonons and the ions.

Let' s study different special cases of the extended {\hub} and see
which models can be recovered.

\vskip4pt plus2pt

{\bf 2.1} {\it Overview of the different models contained in the extended
Hubbard model.}

\vskip4pt plus2pt

In the following table we give a list of different models
contained in the Extended Hubbard Model. A more detailed explanation 
follows in the remarks.

\vskip4pt plus2pt
\vskip4pt plus2pt

{
\item{I)} {Hubbard Model with: $\zeta=0$
\itemitem{a)} generic case:
$\vec \lambda \not=0,\vec \kappa \not=0$
\itemitem{b)} with local phonon interaction: $\vec \lambda \not=0,\vec \kappa=0 
\buildrel{\hbox{\sevenrm Lang-Firsov-transform.}} \over {\Longleftarrow \! = \! = \! = \! = \! \Longrightarrow}$ c)
\itemitem{c)} no local phonon interaction: $\vec \lambda=0, \vec \kappa \not =0
\buildrel{\hbox{\sevenrm Mean field approx.}} \over {= \! = \! = \! \Longrightarrow}$ Standard {\hub}
\itemitem{d)} Standard Hubbard Model plus decoupled Einstein Oscillators: 
$\vec \lambda=0, \vec \kappa =0$
\item{II)} {\hub} with: $\zeta \not=0$, but $\vec \kappa=0$
\itemitem{a)} with local phonon interaction: $\vec \lambda \not=0$
\itemitem{b)} no local phonon interaction: $\vec \lambda=0$ 
\item{III)} General case: $\zeta \not=0, \vec \kappa \not=0, \vec \lambda \not=0$
}

\vskip4pt plus2pt

\remar{Remarks}{

\vskip4pt plus2pt

{
\item{ad I)} When $\vec \kappa=0, \zeta=0$, but $\vec \lambda \not=0$
(case Ib) the Hamiltonian {$\hh$} can be used to describe bipolarons [8],
a model in which there is only the local electron-phonon
interaction. Notice that equivalently, it is possible to
describe the bipolarons with an Hamiltonian of the
type ${\hh}$ with  $\vec \lambda=0$ , but $\vec \kappa \not=0$ (case Ic).
This can be done by performing a unitary transformation,
the Lang-Firsov transformation [9], on
$a_{i \sigma},a_{i \sigma}^{\dagger},p_i,x_i$
with a unitary operator
$$\eqalignno{ &
U(\vec\kappa) \equiv \exp\left(i \vec\kappa \cdot \sum_{l,\sigma} \vec p_l \,
n_{l \sigma}\right). & (7) \cr }
$$
Performing a mean-field approximation on the phonon
variables (when $\vec \lambda=0$), one recovers the standard {\hub}.
\item{ad II)} When $\vec \kappa=0$ the Hamiltonian $\hh$ describes the
{\hub} with phonons added. To see this, remember that the
hopping amplitude in (5) is originally defined by Hubbard [1] as
$$\eqalignno{ 
& T_{ij}=\int d^Dr \,\,\Psi^* (\vec r-\vec R_i-\vec x_i) \left(-{\hbar^2
\nabla_{\vec r}^2 \over 2m}\right)  
\Psi (\vec r-\vec R_j-\vec x_j). & (8) \cr }
$$
where $\Psi(\vec r-\vec R_i)$ is the Wannier electron wave function. 
$T_{ij}$ is a function only of
$\vec a_{ij} \equiv (\vec R_i + \vec x_i) - (\vec R_j + \vec x_j)$.
We approximate the Wannier electron functions with
atomic orbitals, which show an asymptotic exponential decay
$\Psi(\vec r) \sim e^{-\zeta | \vec r | }$ and find 
$$\eqalignno{ 
& \nabla_{\vec a_{ij}} T(\vec a_{ij})= \int d^Dr \,\,\zeta 
{(\vec r-\vec a_{ij}) \over | \vec r-\vec a_{ij} |}\,\, 
\Psi^* (\vec r- \vec a_{ij}) \,\,{\hbar^2 
\nabla_{\vec r}^2 \over 2m}\,\,\Psi(\vec r). & (9) \cr }
$$
Because of the rapid exponential decay of $\Psi(\vec r)$,
we can neglect $\vec r$ in $|\vec r - \vec a_{ij}|$
so that [10]
$$\eqalignno{
& \nabla_{\vec a_{ij}} T(\vec a_{ij})=
-\zeta {\vec a_{ij} \over |\vec a_{ij}|} T(\vec a_{ij}) & (10) \cr }
$$
which integrates to $T(\vec a_{ij})=T_0 e^{-\zeta | \vec a_{ij} |}$.
\item{} $|\vec a_{ij}|=|\vec R_i-\vec R_j+\vec x_i-\vec x_j|$ can be approximated by
$|\vec x_i-\vec x_j| \ll |\vec R_i-\vec R_j|$
such that [5]
$$\eqalignno{
& T_{ij}=t \,\exp\left(-\zeta {(\vec R_i-\vec R_j) \over |\vec R_i-\vec R_j|} 
(\vec x_i-\vec x_j)\right) & (11) \cr }
$$
\item{} with a new constant $t = T_0 \exp(-\zeta|\vec R_i - \vec R_j|)$.}}}

\section{3. Symmetries of the Extended Hubbard Model}

None of the terms added to {\hub} affect the magnetic $SU(2)$-symmetry. 
Hence, in the sequel we shall only be concerned
with the superconductive symmetry of the model. Denote the Jimbo-Drinfel'd
generators of $U_q(su(2))$ with $X^{\pm}$,$H$. They satisfy the following
commutation relations (see e.g. [11]):
$$\eqalignno{ &
\left[ H, X^{(\pm)} \right]= \pm 2 X^{(\pm)}, \qquad
\left[ X^+, X^- \right]={q^{H}-q^{-H} \over q-q^{-1}} & (12) \cr }
$$

\vskip4pt plus2pt

{\bf 3.1} {\it Local commutation relations}

\defin{Definition}{{\it Local representation} of 
the superconductive $U_q(su(2))$ at each site $l$ [4]:
$$\eqalignno{ &
\rho_s(X^+) \equiv \klu =e^{-i \vec{\Phi} 
\cdot \vec p_l} \aalu \aald & (13) \cr
& \rho_s(X^-) \equiv \kld=e^{i \vec{\Phi} 
\vec p_l}\ald \alu =(\klu)^{\dagger} & (14) \cr
& \rho_s(H) \equiv  2 \klz =\nlu + \nld -1
& (15) \cr } $$}

The parameter $\vec{\Phi}$ appearing in (13),(14) does not affect
the $U_q(su(2))$-commutation relations, and for the moment it should
be regarded as a free variable, which will be
determined by the commutation relations with the Hamiltonian.
${\vec \Phi \over 2}$ can be interpreted as the parameter of
a Lang-Firsov transformation [9] of the fermionic
operators.

\vskip4pt plus2pt

\th{Theorem}{1.}{The local part of the Hamiltonian
commutes with the local generators
$$\eqalignno{
& \left[ \klu,H^{(loc)}  \right]= \left[ \kld,H^{(loc)} \right]= 
\left[ \klz,H^{(loc)} \right]=0 & (16) \cr }
$$
if and only if the following conditions are satisfied:
$$\eqalignno{ &
\vec\Phi = {2\vec\lambda \over \hbar M \omega^2}, & (17) \cr
& \mu ={u \over 2} - {1 \over 4} M \omega^2 \hbar^2 \Phi^2 \,\, = \,\, 
{u \over 2} -{\vec{\lambda}^2 \over M\omega^2}. 
& (18) \cr } $$
}

\vskip4pt plus2pt

{\bf 3.2} {\it Global commutation relations}

\vskip4pt plus2pt

Switching signs on $\rho_s(X^{\pm})$ gives again a  representation
of $U_q(su(2))$. It is necessary to consider both representations, i.e.
$$\eqalignno{
& \rho^{\pm}_s(X^{+})=\pm \rho_s(X^{+}), \quad
\rho^{\pm}_s(X^{-})=\pm \rho_s(X^{-}), \quad
\rho^{\pm}_s(H)=\rho_s(H),
& (19) \cr }
$$
{\it if\/} the convention is chosen that fermionic operators at different
sites anticommute.
For each lattice site $l$ a sign $\sigma(l) \in \{ 1,-1 \}$ and the
associated representation $\rho^{\sigma(l)}_s$
will be determined by the symmetry.
The local commutation relations are not affected by this choice.

Further, it is necessary to fix some ordering of the lattice sites 
to be able to define a tensor
product and hence to construct a global symmetry.

\vskip4pt plus2pt

\defin{Definition}{{\it Global representation} of the superconductive $U_q(su(2))$ [4]:

$$\eqalignno{ &
\ku = \bigotimes_{l} \rho_s^{\sigma(l)} (\Delta^{(N-1)} (X^+))
=\sum_l \sigma(l) \prod_{r<l} \, \, e^{\alpha \krz} \,
\,\klu \, \,  \prod_{r>l} \, \, e^{-\alpha^* \krz}, & (20) \cr 
& \kd= \bigotimes_{l} 
\rho_s^{\sigma(l)} (\Delta^{(N-1)} (X^-))
=\sum_l \sigma(l)  \prod_{r<l} \, \, e^{\alpha^* \krz} \, \, \kld \, \, \prod_{r>l}
\, \, e^{-\alpha \krz}, & (21) \cr
& \kz = \bigotimes_{l} \rho_s^{\sigma(l)} (\Delta^{(N-1)} (H))
=\sum_l \klz, & (22) \cr } $$}
$\Delta$ is the coproduct of $U_q(su(2))$ (see e.g. [11]), 
$\alpha=\ln(q)$ is the deformation parameter,
$N$ is the number of lattice sites.

\vskip4pt plus2pt

\th{Theorem}{2.}{The non-local part of  the Hamiltonian
commutes with the global generators 
$$ \eqalignno{
& \left[ \ku,H^{(non-loc)} \right]=
 \left[ \kd,H^{(non-loc)} \right]=\left[ \kz,H^{(non-loc)} \right]=0 
 &  (23) \cr}
$$

\noindent if and only if the following conditions 
are satisfied for $i,j$ nearest neighbours:
$$\eqalignno{
& \sigma(i)=-\sigma(j),  & (24) \cr
& 2 \vec\kappa=\vec\Phi, & (25) \cr 
& Re \alpha =-\vec R_{ij}\cdot \vec\Phi \zeta \hbar, & (26) \cr
& \prod_{i<r<j}e^{\alpha \krz}= 
\prod_{i<r<j}e^{-\alpha^* \krz}. & (27) \cr}
$$}

\vskip4pt plus2pt

\remar{Remarks\ to\ Theorem\ 2:}{Eq. (24) imposes that nearest neighbours
must have opposite signs. This gives a restriction on the possible lattices,
e.g. a triangular lattice is inconsistent with our conventions.

Eq. (25) fixes the parameter $\vec \Phi$ of (13),(14) and relates it to the 
parameter $\vec \kappa$ appearing in the Hamiltonian {$\hh$} (6).
As $\vec \kappa$ can be interpreted as the parameter of a Lang-Firsov
transformation on the Hamiltonian, 
while ${\vec \Phi \over 2}$ can be interpreted as
the parameter of a Lang-Firsov transformation on the generators
of the symmetry, eq. (25) can be interpreted as
a consistency relation, requiring the same transformation 
to be done on the Hamiltonian and on the generators.

Eq. (26) determines the real part of the deformation parameter 
$\alpha$ of the quantum group.
In order for such equation to make sense it is necessary to choose 
a lexicographic
ordering of the lattice sites, so that the sign of the term 
$\vec R_{ij}$  does not
depend on the particular couple $i,j$. 
Notice that there is no restriction on the imaginary
part of the deformation parameter; we can safely choose it to be real.

Eq. (27) is trivially satisfied if $\alpha = 0$; it
is a strong condition in the case $\alpha \not= 0$:
It then implies for nearest neighbours $i < j$ that there cannot be
a site $r$ in between them, {\it i.e.\/} with $i < r < j$.
However, such a condition requires that the lattice
on which the {\hub} is defined is one-dimensional,
and that a ordering of sites is chosen in which
the sites are numbered from left to right in increasing {\it or\/}
decreasing order.}

\vskip4pt plus2pt

\Proof  (Sketch) The (undeformed) generator
$\kz$ commutes with $H^{(non-loc)}$ given by
(5) as in the classical ($q = 1$) case.

We still have to calculate
$$ \eqalignno{
& \left[ \ku,H^{(non-loc)} \right]= &  \cr 
& t {\displaystyle \sum_{\la i<j \ra}} e^{-\hbar 
\zeta \vec R_{ij} \cdot \vec \kappa}
\Bigg\{ 
(\aaid \aaiu \aajd \aaju  \aiu \ajd
-\aaid \aaiu \aajd \aaju \aid \aju) & \cr
& \times \Bigg[ Z_{ij} \left( 2 \cosh \left({1 \over 2} \vec R_{ij}
\cdot \vec{\Phi} \zeta \hbar \right) 
-  2 \cosh \left({1 \over 2}   \vec R_{ij} \cdot
\vec{\Phi} \zeta \hbar + {1 \over 2} \alpha^* \right) \right) & \cr
& + Z_{ji} \left(  2 \cosh \left( {1 \over 2} \vec R_{ij} \cdot
\vec{\Phi} \zeta \hbar \right)
-  2 \cosh \left({1 \over 2}   \vec R_{ij} \cdot
\vec{\Phi} \zeta \hbar + {1 \over 2} \alpha \right) 
\right) \Bigg] & \cr
& + (\aaiu \aajd \aaju \aju + \aaid \aajd \aaju \ajd)
\epi \Bigg[ Z_{ij} (\eajim -1)  + Z_{ji}
\left( e^{-\vec R_{ij} \cdot  
\vec{\Phi} \zeta \hbar} \eaijm  -1 \right) \Bigg] & \cr 
& +(\aaid \aaiu \aajd \aid + \aaid \aaiu \aaju \aiu) \epj
\Bigg[  Z_{ij} (e^{\vec R_{ij} \cdot  \vec{\Phi} \zeta \hbar}
\eajim - 1) + Z_{ji} (\eaijm-1) \Bigg] & \cr
& +   (\aaid \aaju -\aaiu \aajd) \Bigg[ Z_{ij} \epi \eajim 
+ Z_{ji} \epj \eaijm ) \Bigg] \Bigg\} & \cr
& +   {\displaystyle \sum_l \sum_{\la i,j \ra ,i<l<j}} 
\sigma (l) e^{-\hbar \zeta \vec R_{ij} 
\cdot \vec \kappa} e^{-i \vec{\Phi} \cdot \vec p_l} \aalu \aald 
{\displaystyle
\prod_{r<l,r \not=i} e^{\alpha \krz}} 
{\displaystyle \prod_{r>l,r \not=j}} e^{-\alpha^* \krz} & \cr 
& \times   \left[ e^{\alpha \kiz} e^{-\alpha^* \kjz},
e^{i \vec{\kappa} \cdot  (\vec p_i-\vec p_j)
+\zeta \vec R_{ij} \cdot (\vec x_i-\vec x_j)} \aaju \aiu 
+\aajd \aid +h.c. \right], & \cr }
$$
where we have introduced the abbreviation
$$\eqalignno{
& Z_{ij}=\sigma(i) \ei \prod_{r<i, r \not= j}e^{\alpha \krz} 
\prod_{r>i, r \not= j}e^{-\alpha^* \krz}. & (28) \cr }
$$

The sums containing different numbers of fermionic operators
are linearly independent and
must all vanish separately. (We have fixed a normal ordering for the
fermionic operators.)
Studying  the term containing $\aaid \aaju -\aaiu \aajd$,
the conditions (24)-(27) are obtained. It turns out that the same conditions
guarantee that also the other sums vanish.

\vskip4pt plus2pt

{\bf 3.3} {\it Summary}

\vskip4pt plus2pt

{\offinterlineskip \tabskip=0pt
\halign {\strut  \vrule# & \quad # & \vrule# & \quad # & \vrule# \cr
\noalign{\hrule}
&&&& \cr
\noalign{\vskip-5pt}
& {\bf Classical Symmetry} \hfill && {\bf Quantum Symmetry} \hfill & \cr 
& $\alpha=0$ \hfill && $\alpha \not=0$, $q=\exp(\alpha)$ \hfill & \cr
& $D$ arbitrary \hfill && $D=1$ \hfill & \cr
&&&& \cr
\noalign{\vskip-5pt}
\noalign{\hrule}
&&&& \cr
\noalign{\vskip-5pt}
& Id), IIb): $SU(2)_s$-symmetry; $\mu={u \over 2}$ &&
III): $SU_q(2)_s$-symmetry; & \cr
&&& $Re \alpha = 2 \kappa \zeta \hbar$,  \hfill & \cr
& Ia):  $SU(2)_s$-symmetry; \hfill && $\mu=u/2-\hbar^2 \kappa^2 M \omega^2$,
$\vec \lambda=\hbar M \omega^2 \vec \kappa$ \hfill
& \cr
& $\mu=u/2-\hbar^2  \vec \kappa^2 M \omega^2$,
$\vec \lambda= \hbar M \omega^2 \vec \kappa$ \hfill &&
& \cr
&&&& \cr
& Ib),Ic),IIa): no symmetry  \hfill &&& \cr
&&&& \cr
\noalign{\hrule}}}

\remar{Remark:}{ After choosing an appropriate ordering
of the lattice sites for model III) in the 1-dimensional case,
we have set $\kappa\equiv -\vec R_{ij} \cdot \vec \kappa$.
The models Ib),Ic) and IIa) have no symmetry,
because the condition
$2 \vec \kappa={2 \vec \lambda \over \hbar M \omega^2}$ is not
satisfied.}

\section{4. Discussion} 

The standard {\hub} Id) (with decoupled Einstein oscillators) and the
{\hub} IIb) without local phonon interaction ($\vec \lambda=0$) 
have a superconductive $SU(2)_s$ symmetry at ``half-filling'' $\mu=u/2$.
Two other models Ia), III) with local phonon interaction ($\lambda \not=0$)
have superconducting symmetries $SU(2)_s$ and $SU_q(2)_s$ respectively
at $\mu=u/2-\hbar^2 \kappa^2 M \omega^2$.
It is well known [9] that a Lang-Firsov transformation can be
performed on the model Ia), shifting the parameters and thereby
eliminating the local phonon interaction $\vec \lambda$.
The resulting Hamiltonian has a $SU(2)_s$-symmetry at 
``half-filling'' $\mu=u/2$. It turns out that a Lang-Firsov transformation
with the unitary operator (7) can also be used to eliminate the local
electron-phonon interaction term from the Extended Hubbard Model.
The parameters are shifted according to
$$ \eqalignno{
& \vec\lambda \rightarrow  \vec\lambda - M \omega^2 \hbar \vec\kappa, \,
u \rightarrow u -2 \hbar \vec\lambda\cdot\vec\kappa + M \omega^2 \hbar^2 \vec \kappa^2, \,
\mu \rightarrow \mu + \hbar \vec\lambda\cdot\vec\kappa - 1/2 
M \omega^2 \hbar^2 \vec \kappa^2 & (29) \cr}
$$
and the hopping term $H^{(non-loc)}_{el-ph}$ becomes
$$ \eqalignno{
& t \sum_{\langle i<j\rangle,\sigma }  e^{\zeta \vec R_{ij} 
\cdot (\vec x_i- \vec x_j)} q^{1\over 2} \,\, \aajs \ais \Big(1 + (q^{-{1\over 2}}- 1) n_{i, -\sigma} \Big)
\Big(1 + (q^{1\over 2}- 1)n_{j, -\sigma} \Big)  \, + \, h.c. & (30)}
$$
The resulting Hamiltonian has a superconducting
$SU_q(2)_s$ quantum symmetry at ``half filling''
$\mu=u/2$ (!) with $\vec \lambda=0$ and only for $D=1$.

It is even possible to eliminate all phonon terms from
this Quantum Symmetric Hamiltonian by a mean field
approximation [8,9] {\it without breaking the quantum
symmetry}. More precisely one performs a thermal average
over  $H_{ph}$-eigenstates and assumes uncorrelated Einstein oscillators
(note that $\vec \lambda=0$ !). The exponential in the hopping term
is then approximated by a temperature-dependent constant.

{\bf Acknowledgements:}\ BLC would like to thank the organizers of the Minisemester on
Quantum Groups at the Banach Center and the Forschungszentrum
Karlsruhe Technik und Umwelt for financial support.

\references{Nov}{
\item{[1]} J.~Hubbard, {\it Proc.\ R.\ Soc.\ London\/} {\bf A 276}
(1963) 238
\item{[2]}C.~N.~Yang, S.~C.~Zhang, {\it Mod.\ Phys.\ Lett.\/}
{\bf B 4} (1990) 759
\item{[3]}C.~N.~Yang, {\it Phys.\ Rev.\ Lett.\/} {\bf 63} (1989) 2144
\item{[4]}A.~Montorsi, M.~Rasetti, {\it Phys.\ Rev.\ Lett.\/} 
{\bf 72} (1994) 1730
\item{[5]} B.~L.~Cerchiai, P.~Schupp, {\it 
J.\ Phys.\/} {\bf A 29} (1996) 845
\item{[6]} T.~Holstein, {\it Ann.\ Phys.\/} {\bf 8} (1959) 325
\item{[7]} A.~L.~Fetter, J.~D.~Walecka, 
{\it Quantum Theory of Many Particle
Systems}, (McGraw Hill, New York, 1971) p.\ 396
\item{[8]} S.~Robaszkiewicz, R.~Micnas, J.~Ranninger, {\it Phys.\ Rev.\/}
{\bf B 36} (1987) 180 
\item{[9]} I.~G.~Lang, Y.~A.~Firsov, {\it Sov.\ Phys.\ JETP\/}
{\bf 16} (1963) 1301
\item{[10]} S.~Bari\u{s}i\'{c}, J.~Labb\'{e}, J.~Friedel, 
{\it Phys.\ Rev.\ Lett.\/}
{\bf 25} (1970) 919
\item{[11]} M.~Jimbo, {\it Lett.\ Math.\ Phys.\/} 
{\bf 11} (1986) 247
}
\bye